\documentclass[preprint, showpacs, floatfix, nofootinbib]{revtex4}
\usepackage{graphicx}
\usepackage{amsmath}

\begin{document}

\title{Evolution of elliptic and triangular flow as a function of $\sqrt{s_{NN}}$ 
in a hybrid model}

\author{Jussi Auvinen}
\email{auvinen@fias.uni-frankfurt.de}
\affiliation{Frankfurt Institute for Advanced Studies (FIAS), Ruth-Moufang-Strasse 1, 
D-60438 Frankfurt am Main, Germany}
\author{Hannah Petersen}
\email{petersen@fias.uni-frankfurt.de}
\affiliation{Frankfurt Institute for Advanced Studies (FIAS), Ruth-Moufang-Strasse 1, 
D-60438 Frankfurt am Main, Germany}
\affiliation{Institut f\"ur Theoretische Physik, Goethe Universit\"at, Max-von-Laue-Strasse 1, 60438 Frankfurt am Main, Germany}
\begin{abstract}
We study the collision energy dependence of elliptic flow $v_2$ and triangular flow $v_3$ 
in Au+Au collisions within the energy range $\sqrt{s_{NN}}=5-200$ GeV, utilizing a 
transport + hydrodynamics hybrid model. The transport part is described by 
the Ultrarelativistic Quantum Molecular Dynamics (UrQMD) approach, combined with an 
intermediate (3+1)-dimensional ideal hydrodynamical evolution phase using a chiral model 
equation of state. We find the decrease of $v_2$ produced by hydrodynamics at lower 
collision energies partially compensated by the transport dynamics. This does not apply 
to $v_3$, which falls to 0 in midcentral collisions at $\sqrt{s_{NN}}=5$ GeV. We conclude 
that the triangular flow provides the clearer signal for the formation of low-viscous fluid 
in heavy ion collisions.

\end{abstract}

\pacs{24.10.Lx,24.10.Nz,25.75.Ld}

\maketitle

\section{Introduction}

Several lattice calculations \cite{Fodor:2004nz,Ejiri:2008xt,Gavai:2008zr} have predicted 
the existence of a critical point in the QCD phase diagram, which marks the boundary of 
cross-over and first-order phase transition between the hadronic and QCD matter in the plane of 
baryochemical potential $\mu_B$ and temperature $T$. However, only the cross-over phase transition 
was seen in the continuum extrapolated results \cite{Aoki:2006we,Endrodi:2011gv}. 
In 2010, a beam energy scan program was launched at the Relativistic Heavy Ion Collider 
(RHIC) to study the features of the phase diagram and to investigate if any signals
of a critical point can be found within the available range of $\mu_B$.

By running experiments at different beam energies, collisions with varying peak temperatures 
and values of $\mu_B$ are produced that span a large region in the phase diagram of strongly 
interacting matter. So far, the scanned energy range reaches from $\sqrt{s_{NN}}=200$ GeV 
down to 7.7 GeV, extending the baryochemical range from $\mu_B \sim 0$ up to $\sim 400$ MeV. 
Additional fixed target collisions down to $\sim 3$ GeV have been planned, which increase 
the $\mu_B$-range even further \cite{Odyniec:2013kna}. This region of the phase diagram 
will also be the target for more detailed studies with much higher luminosities at 
the Facility for Antiproton and Ion Research (FAIR), which is currently under construction.

One of the key observables considered as the evidence of the formation of the strongly 
interacting matter, ``quark-gluon plasma'' (QGP) at RHIC and the Large Hadron Collider 
(LHC) is the elliptic flow, typically characterized by coefficient $v_2$ of 
the Fourier expansion of the azimuthal angle distribution of the final state particles. 
One would expect $v_2$ to decrease at lower beam energies, as the duration of the QGP phase 
gets smaller. However, the inclusive charged hadron $v_2$ has demonstrated surprisingly weak 
dependence on the collision energy between 7.7 and 39 GeV \cite{Adamczyk:2012ku}. 
On the other hand, the preliminary results for the next Fourier coefficient $v_3$, known as 
triangular flow, display a clearer change in magnitude for this observable between 
$\sqrt{s_{NN}}=27 - 200$ GeV \cite{Pandit:QM2012}. 
The weak energy dependence of the elliptic flow thus requires an explanation.

The beam energy dependence of the collective flow has been recently studied with several 
different models \cite{Konchakovski:2012yg,Shen:2012vn,Solanki:2012ne,Plumari:2013bga,
Lacey:2013qua,Denicol:2013nua}. 
The method chosen for the present study is a hybrid approach, where a transport model 
-- a microscopic description of the system -- is utilized for the non-equilibrium phases at 
the beginning and in the end of a heavy-ion collision event, while a macroscopic 
hydrodynamical description is used to model the hot and dense intermediate stage 
incorporating the phase transition between the QGP and hadronic matter. Such a hybrid model 
provides a consistent framework for investigating both high-energy heavy ion collisions 
with negligible net-baryon density and a large hydrodynamically evolving medium, 
and the collisions at smaller energies with finite net-baryon density, where 
no such medium is formed. Thus, this approach is optimal for studying the beam energy 
dependence of the elliptic and triangular flow.

The next Section provides a brief account of the main features of the applied hybrid model.
The simulation results are presented in Section~\ref{sec_results}. 
After comparing the transverse mass spectra 
of various particles against the experimental data in subsection~\ref{sec_res_spectra}, we 
study the elliptic and triangular flow in subsections~\ref{sec_res_v2} and 
\ref{sec_res_v3}, respectively, concluding with an investigation of 
the dependence of flow coefficients on the initial collision geometry 
in~\ref{sec_res_geometry}. We then summarize our findings in Section~\ref{sec_summary}.

\section{Hybrid model}
\label{sec_model}

This study was performed using a Boltzmann + hydrodynamics hybrid model described 
in \cite{Petersen:2008dd}. In this framework, both the initial state before equilibrium 
and the final state with hadronic rescatterings and decays is calculated within the 
Ultrarelativistic Quantum Molecular Dynamics (UrQMD) string / hadronic cascade 
\cite{Bass:1998ca,Bleicher:1999xi}.

The intermediate hydrodynamical evolution starts, when the two colliding nuclei have passed 
through each other:
\begin{equation}
\label{eq_tstart}
t_{\mbox{start}}=\mbox{max}\{\frac{2R}{\sqrt{\gamma_{CM}^2-1}},0.5 \mbox{ fm}\},
\end{equation}
where $R$ represents the nuclear radius and $\gamma_{CM}=\frac{1}{\sqrt{1-v_{CM}^2}}$ 
is the Lorentz factor in the center-of-mass frame of the colliding nuclei. A minimum 
starting time of 0.5 fm is chosen based on the hybrid model results at the collision energy 
$\sqrt{s_{NN}}=200$ GeV \cite{Petersen:2010zt}.
At this time, the energy-, momentum- and baryon number densities of the particles are 
mapped onto the hydro grid. The particles are represented by 3D Gaussian distributions 
that are Lorentz-contracted in the beam direction. The width parameter of these Gaussians 
is chosen to have the value $\sigma=1.0$ fm to preserve the event-by-event initial state 
fluctuations. The spectator particles, which do not participate on the hydrodynamical 
evolution, are propagated separately in the cascade.

The evolution of the system in the intermediate phase is based on (3+1)-D ideal 
hydrodynamics, solving the evolution equations using the SHASTA algorithm 
\cite{Rischke:1995ir,Rischke:1995mt}. The equation of state (EoS) is based on a 
hadronic chiral parity doublet model including quark degrees of freedom and the thermal 
contribution of the Polyakov loop \cite{Steinheimer:2009nn,Steinheimer:2011ea}. This EoS 
qualitatively agrees with the lattice QCD data at $\mu_B=0$ and, most importantly, is also 
applicable at finite baryon densities. After the last step of the hydrodynamical evolution, 
the active equation of state is changed from the deconfinement EoS to the hadron gas EoS, 
to ensure that the active degrees of freedom on both sides of the transition hypersurface 
are exactly equivalent \cite{Steinheimer:2009nn}.

The transition from hydro to transport, also known as ``particlization'', is done when 
the energy density $\epsilon$ is smaller than the critical value $2\epsilon_0$, 
where $\epsilon_0=146$ MeV/fm$^3$ represents the nuclear ground state energy 
density. This corresponds roughly to a switching temperature $T \approx 154$ MeV at 
$\sqrt{s_{NN}}=200$ GeV Au+Au collisions \cite{Huovinen:2012is}.  
While the switching criterion with respect to the energy density is kept constant over all 
collision energies, it will correspond to different combinations of temperature and 
baryochemical potential at different values of $\sqrt{s_{NN}}$.

The four-dimensional iso-energy density spacetime surface is constructed using 
the Cornelius hypersurface finder \cite{Huovinen:2012is}. From this hypersurface, 
the particle distributions are generated according to the Cooper-Frye 
formula. Rescatterings and final decays of these particles are then computed in UrQMD. 
The end result is a distribution of particles which is directly comparable against 
the experimental data.

The dynamic change in the importance of the non-equilibrium transport and the hydrodynamic 
part of the evolution and having a proper equation of state applicable at high net baryon 
densities are the main advantages of this hybrid approach. As it is enough for the purposes 
of this study to reach a qualitative agreement with the experimental results, we neglect 
the viscosity effects during the hydrodynamical evolution. The high viscosity in 
the hadron gas phase is included, however. 

Compared to the previous investigations of the elliptic flow using the same hybrid 
approach \cite{Petersen:2009vx,Petersen:2010md}, the new features in this study are the new 
implementation of the Cooper-Frye hypersurface finder and particlization, described above,
and replacing the reaction plane (RP) analysis with the event plane (EP) method 
\cite{Poskanzer:1998yz,Ollitrault:1997di} when computing $v_2$ and $v_3$ from 
the particle momentum distributions.  

\section{Results}
\label{sec_results}

\subsection{Particle spectra}
\label{sec_res_spectra}

\begin{figure}
\centering
\includegraphics[width=7cm]{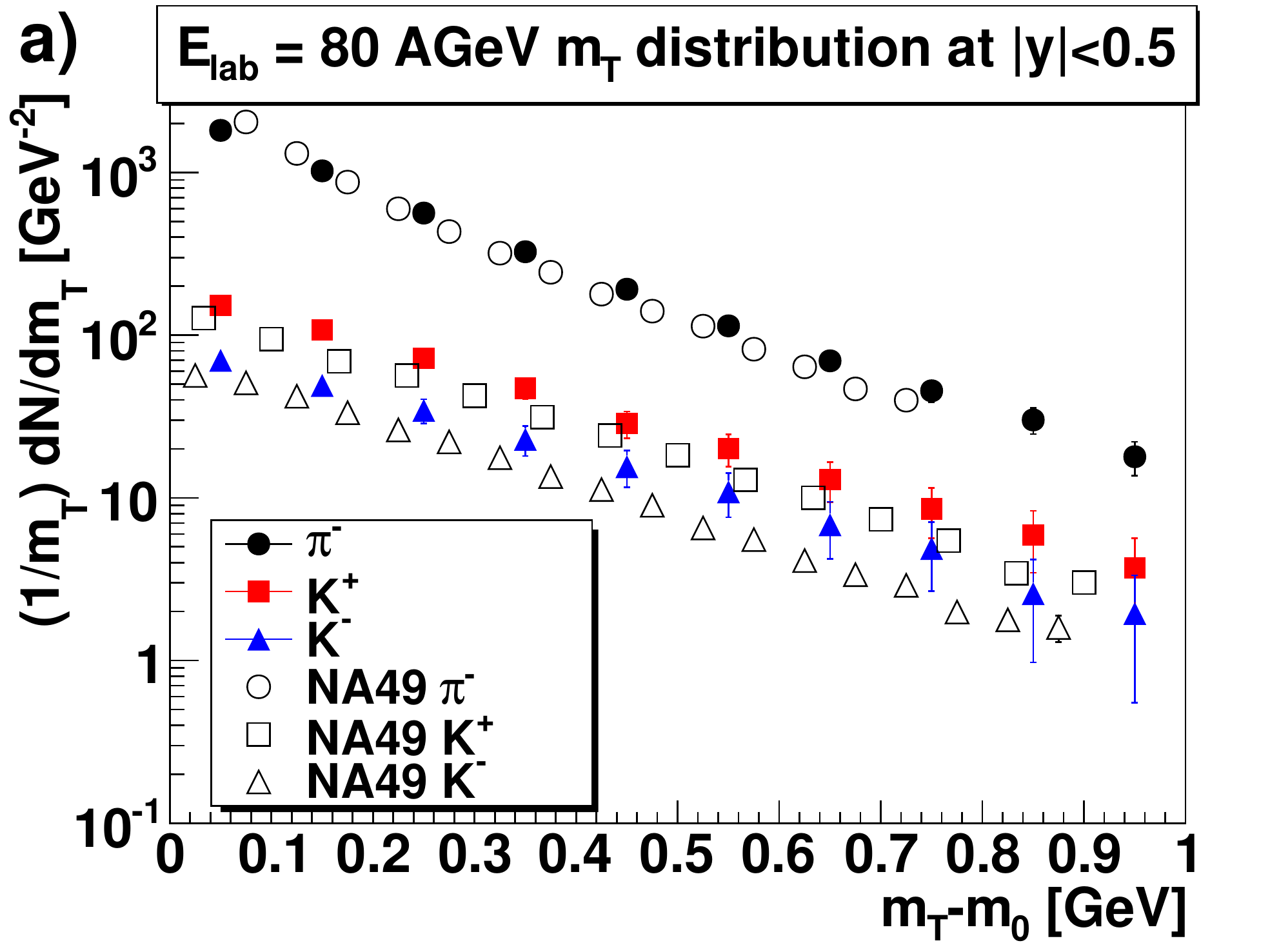}
\includegraphics[width=7cm]{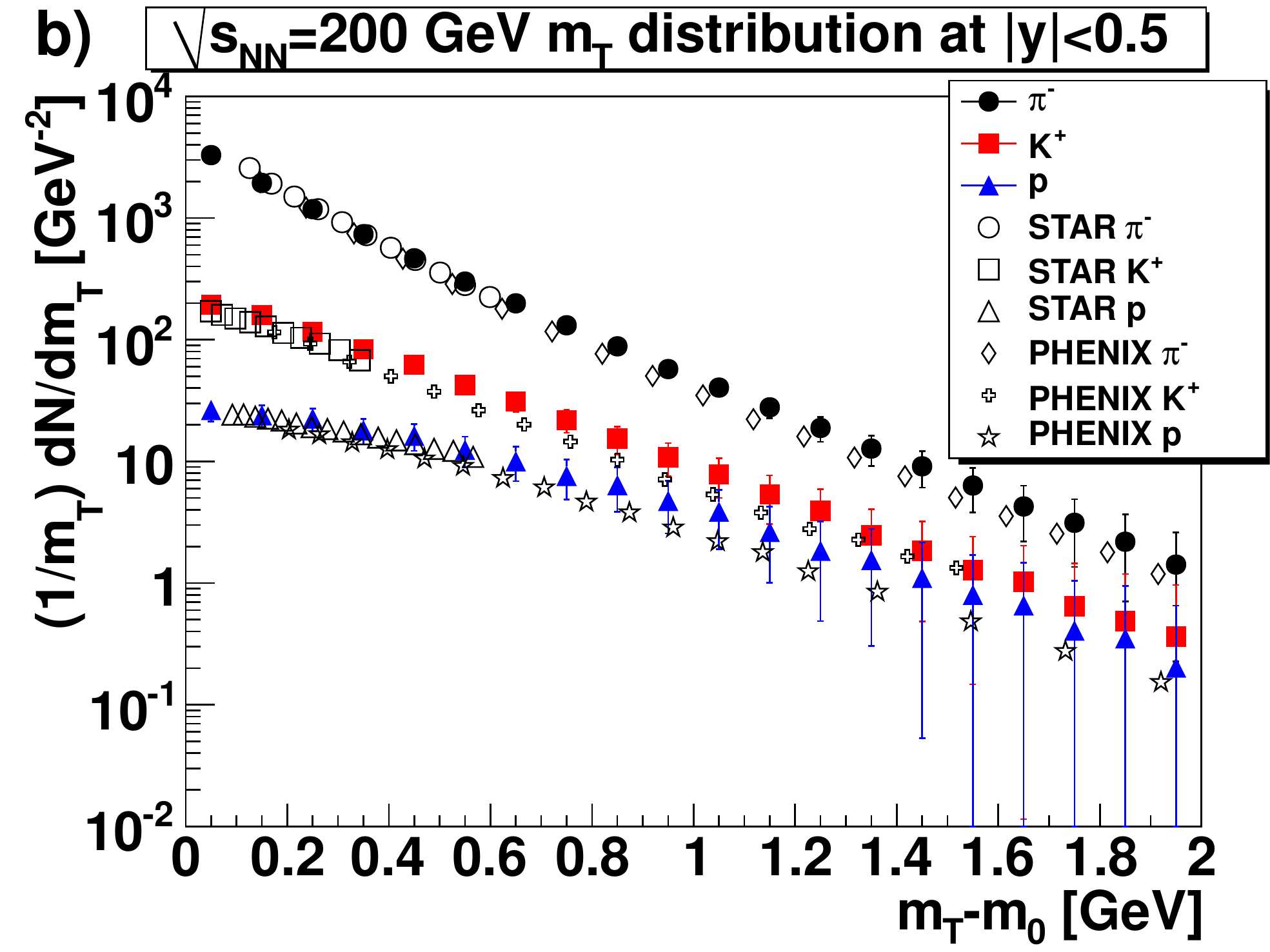}
\caption{(Color online) Transverse mass $m_T$ -spectra at midrapidity $|y|<0.5$.  
a) $m_T$-spectra for $\pi^-, K^+$ and $K^-$ in Pb+Pb -collisions with impact parameter 
$b = 0 - 4$ fm, compared to the NA49 data \cite{Afanasiev:2002mx} at beam energy 
$E_{\textrm{lab}}=80$ AGeV.  
b) $m_T$-spectra for $\pi^-, K^+$ and $p$ in $\sqrt{s_{NN}}=200$ GeV Au+Au -collisions with 
impact parameter $b = 0 - 3.4$ fm, compared to the 0-5\% centrality data from STAR 
\cite{Adams:2003xp} and PHENIX \cite{Adler:2003au}.}
\label{Figure_mt_spectra}
\end{figure}

Our first step is to check how well the current setup of the hybrid model reproduces 
experimental data for bulk observables. The $m_T$ spectra at midrapidity $|y|<0.5$ for $\pi^-, K^+$ and $K^-$ 
in Pb+Pb -collisions with beam energy $E_{\textrm{lab}}=80$ AGeV
(corresponding to the collision energy $\sqrt{s_{NN}}\approx 12$ GeV) is illustrated in 
Figure~\ref{Figure_mt_spectra}a. A good agreement with the NA49 data 
\cite{Afanasiev:2002mx} is found, although the pion slope is a little too flat and there is 
an excess of kaons produced. Similar results are found for $\pi^-, K^+$ and $p$ in 
$\sqrt{s_{NN}}=200$ GeV Au+Au -collisions, as shown in Figure~\ref{Figure_mt_spectra}b.
For the purpose of the current investigation the agreement with the experimental data is 
sufficient; for future studies the particlization energy density value can be adjusted to 
achieve a better agreement with the measured spectra. 

\subsection{Elliptic flow}
\label{sec_res_v2}

\begin{figure}
\centering
\includegraphics[width=8cm]{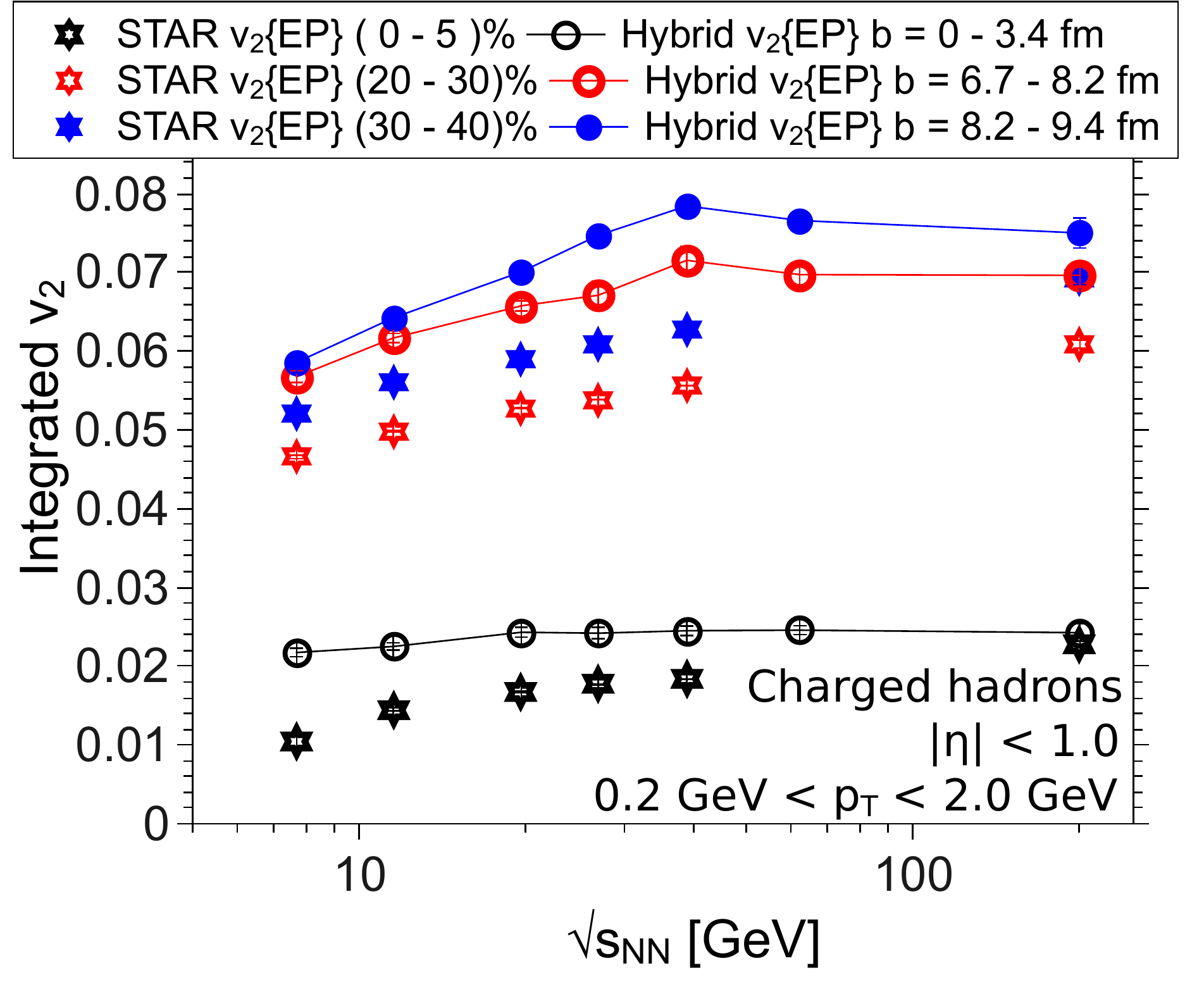}	
\caption{(Color online) Integrated elliptic flow $v_2\{\textrm{EP}\}$ for charged hadrons 
with $0.2 < p_T < 2.0$ at midrapidity $|\eta|<1.0$ in Au+Au -collisions, 
for collision energies $\sqrt{s_{NN}}=7.7 - 200$ GeV and three different impact parameter 
ranges, compared with the STAR data \cite{Adamczyk:2012ku,Adams:2004bi}.}
\label{Figure_v2_star}
\end{figure}

Figure~\ref{Figure_v2_star} shows the hybrid model result for the integrated elliptic flow 
$v_2\{\textrm{EP}\}$ for charged particles with transverse momentum 0.2 GeV $< p_T < 2.0$ 
GeV produced in Au+Au -collisions in $|\eta|<1.0$ pseudorapidity, 
compared with the STAR data for three centrality classes: 
(0-5)\%, (20-30)\% and (30-40)\%. 
These centralities are respectively represented by the impact parameter intervals 
$b = 0-3.4$ fm, $b = 6.7-8.2$ fm and $b = 8.2-9.4$ fm in the model, where the choice 
of values is based on the optical Glauber model estimates \cite{Eskola:1988yh,Miskowiec-web}. 

\begin{figure}
\centering
\includegraphics[width=4.7cm]{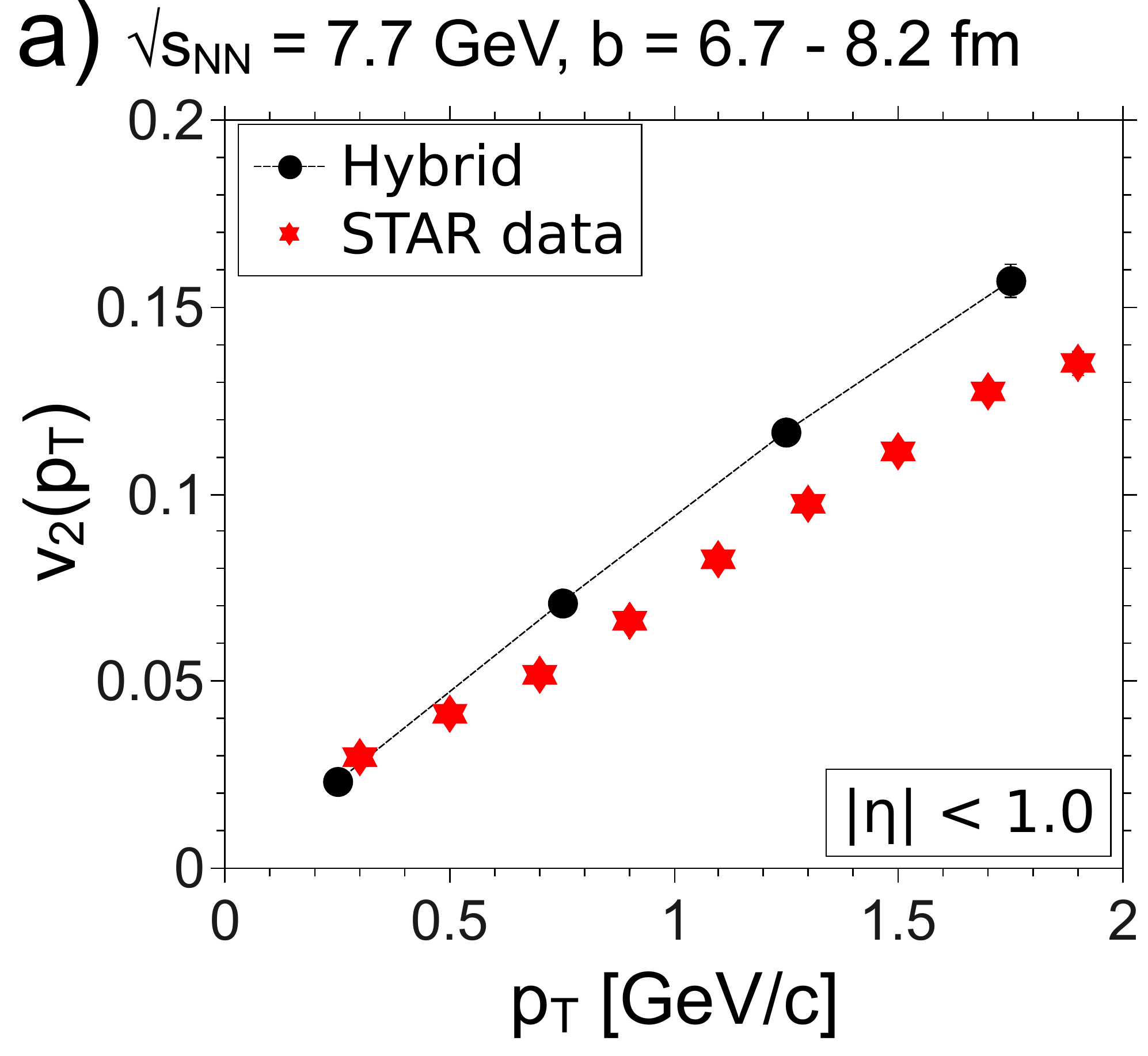}
\includegraphics[width=4.7cm]{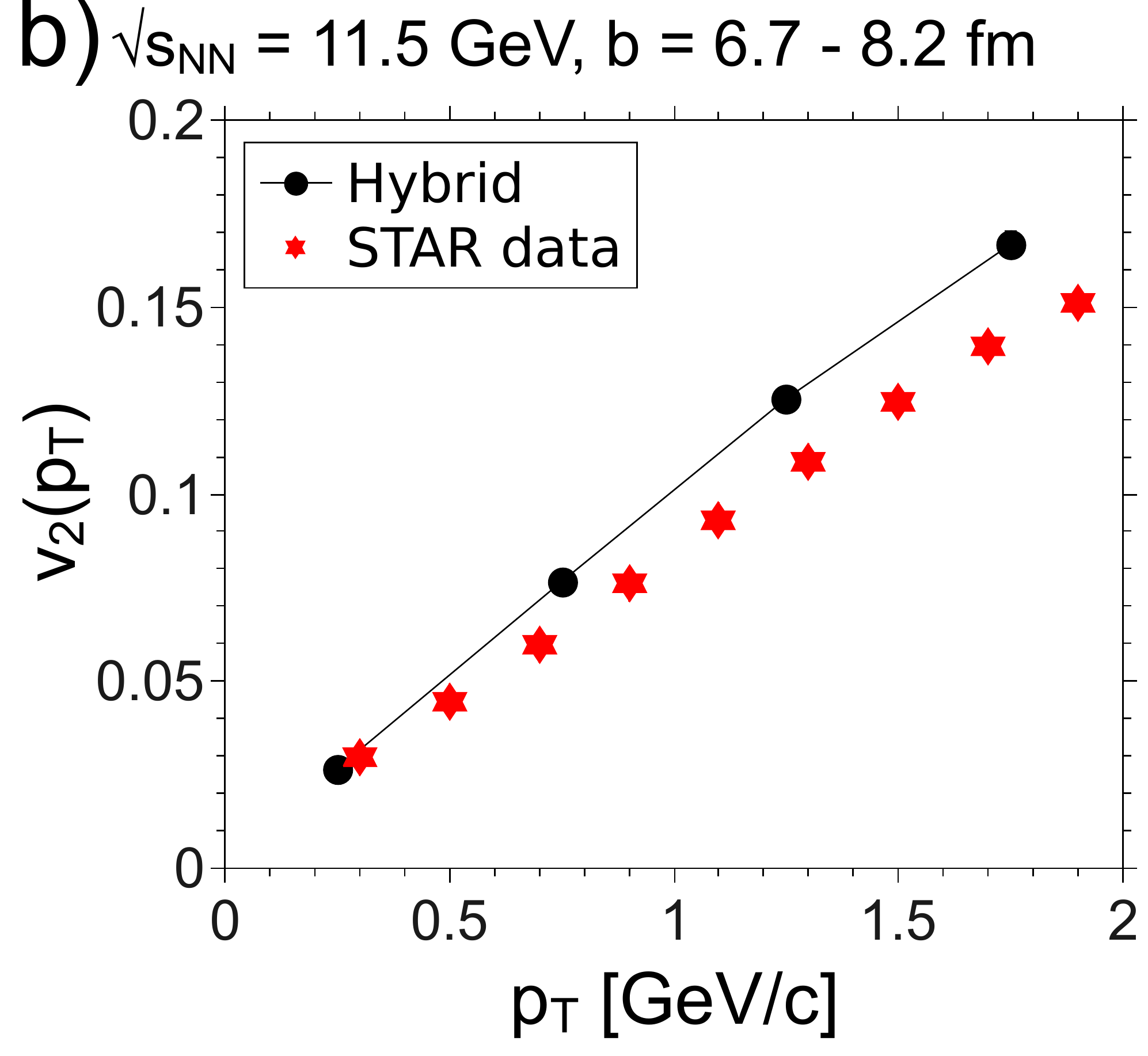}
\includegraphics[width=4.7cm]{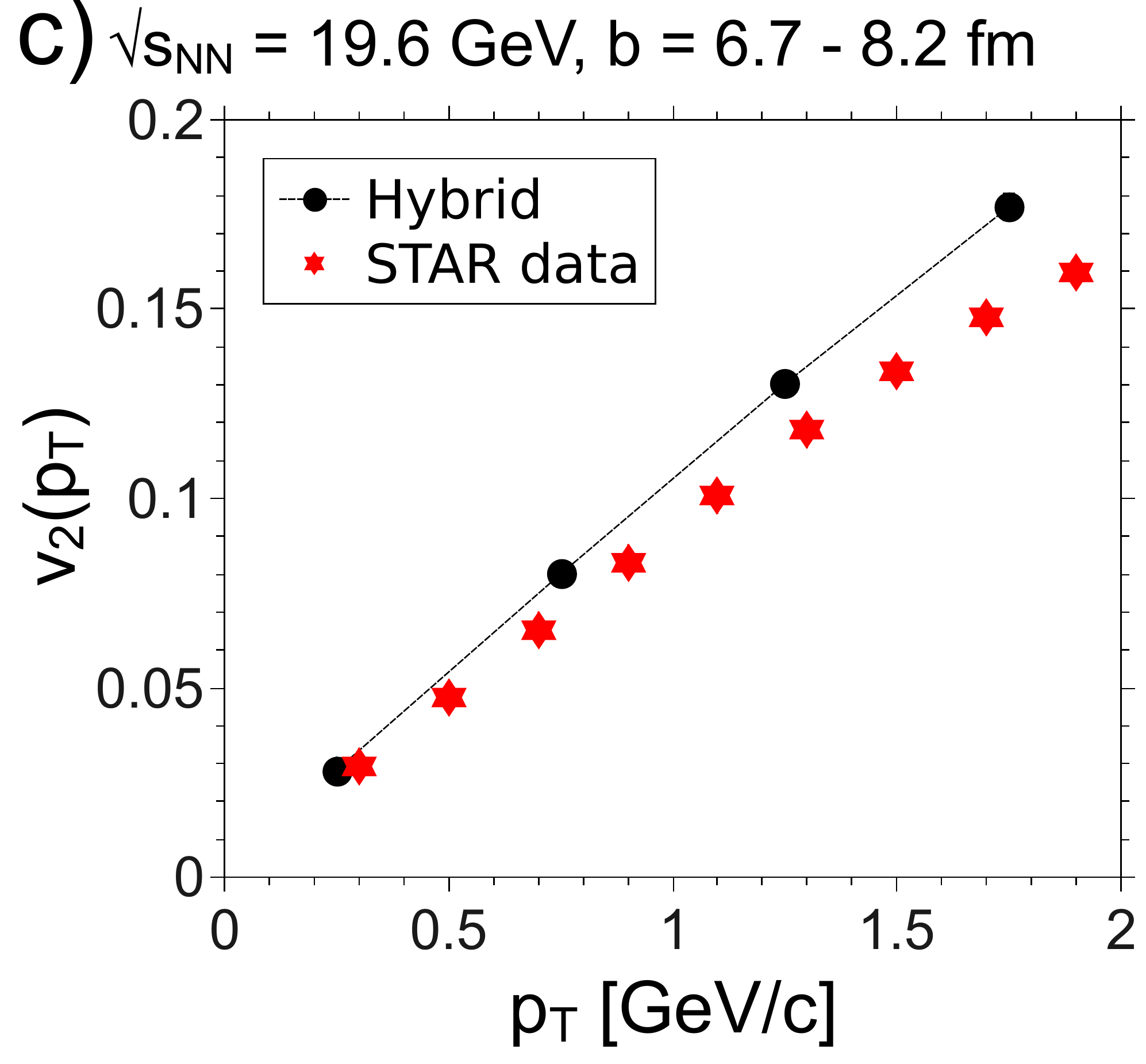}	
\includegraphics[width=4.7cm]{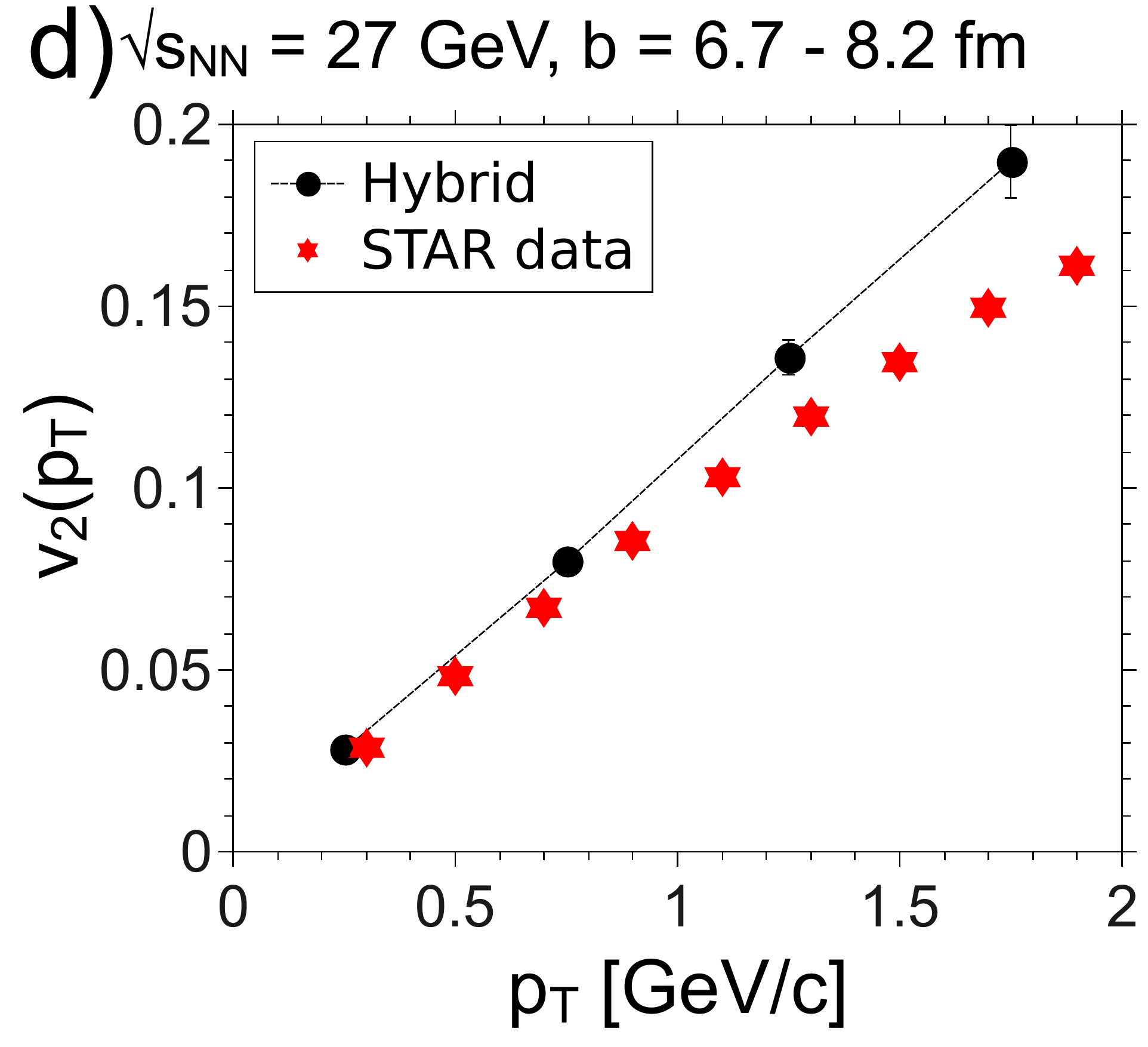}
\includegraphics[width=4.7cm]{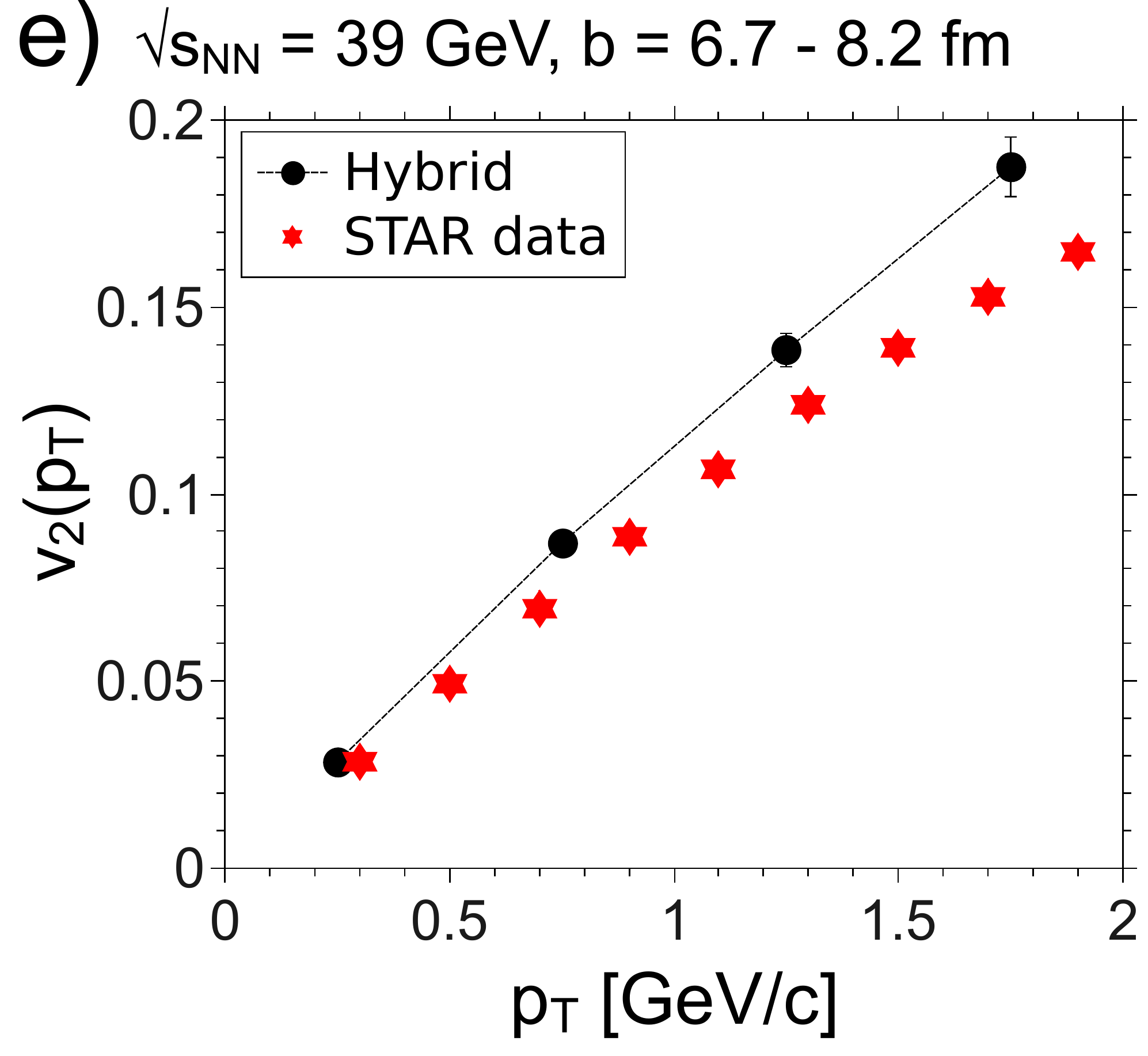}
\caption{(Color online) (a-e) Differential $v_2\{\textrm{EP}\}$ at midrapidity $|\eta|<1.0$ 
for collision energies $\sqrt{s_{NN}}=7.7 - 39$ GeV in impact parameter range $b = 6.7-8.2$ 
fm, compared with the STAR data in (20-30)\% centrality \cite{Adamczyk:2012ku}.}
\label{Figure_v2_pt_star}
\end{figure}

With the chosen parameters, the hybrid model systematically overshoots the experimental 
data. The examination of the $v_2(p_T)$ produced by the simulations 
(Figure~\ref{Figure_v2_pt_star}) reveals that the overshoot is worse at higher $p_T$, 
while at the lower limit of the $p_T$-cut the produced flow agrees with the data. It remains 
as a question for a future study to see if both the particle spectra and the flow can be 
made to match the data with the same choice of parameters; likely the viscous corrections 
will prove to be necessary. However, for this investigation the most important thing is the 
qualitative agreement with the data -- for midcentral collisions, the observed modest 
collision energy dependence is well reproduced by the model. In the most central collisions, 
the elliptic flow energy dependence is even weaker than in experiments, to the point of 
being almost constant.

\begin{figure}
\centering
\includegraphics[width=7.6cm]{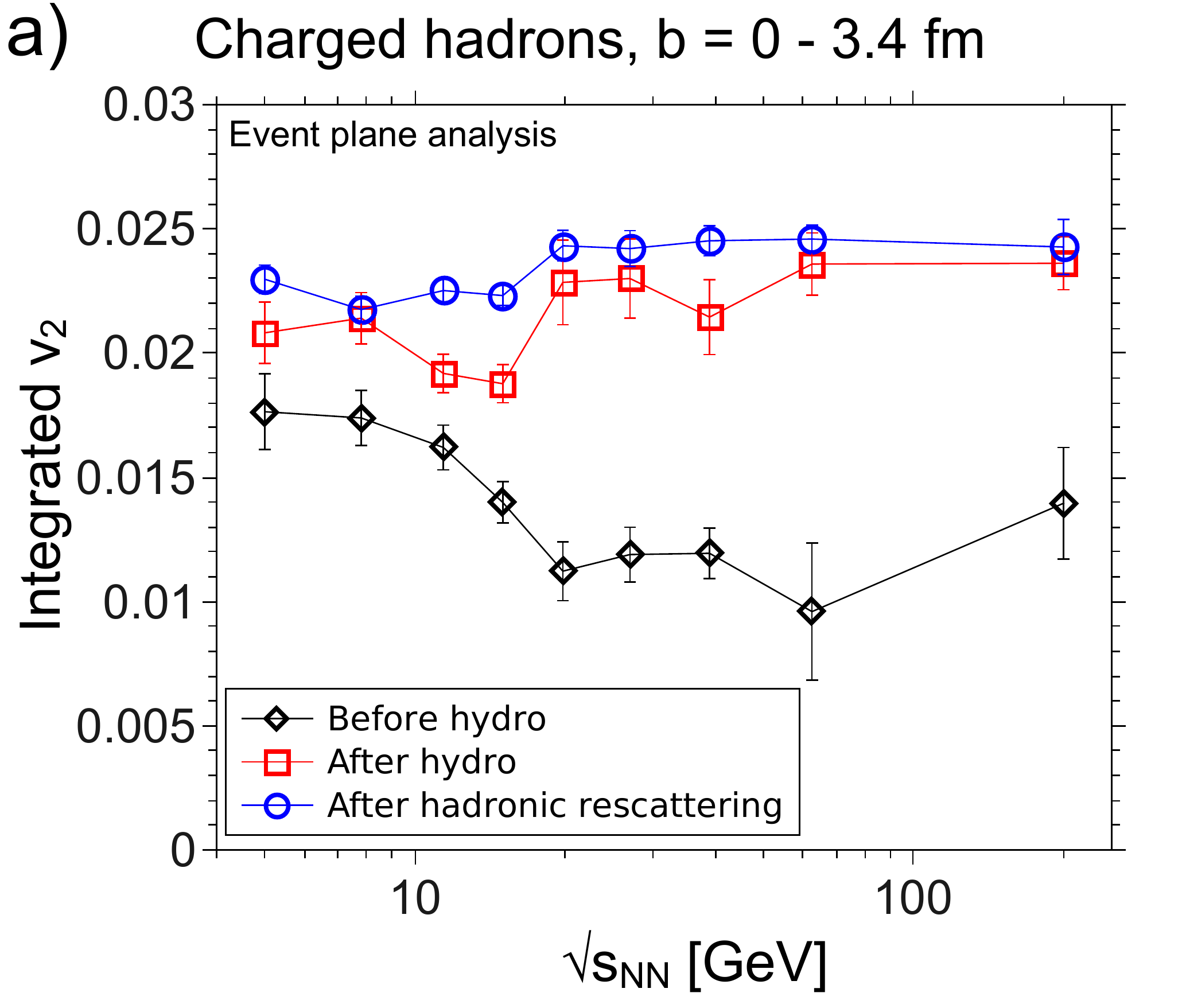}
\includegraphics[width=7.3cm]{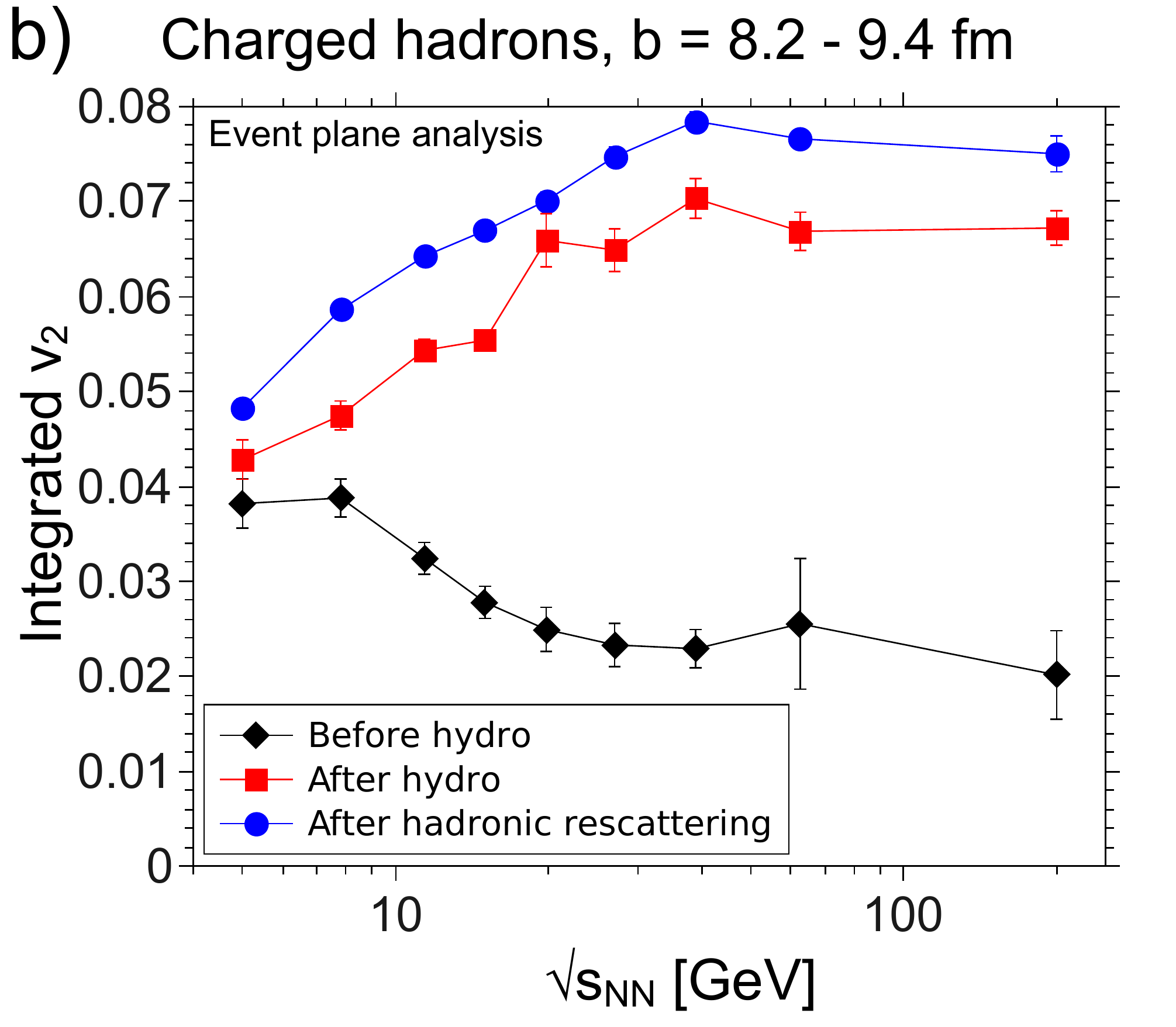}
\caption{(Color online) Magnitude of $v_2\{\textrm{EP}\}$ at the beginning of 
hydrodynamical evolution (diamonds), immediately after particlization (squares) and after 
the full simulation (circles, the same as in Fig.~\protect\ref{Figure_v2_star}) at 
a) central collisions and b) midcentral collisions.}
\label{Figure_v2_phases}
\end{figure}

In order to understand why the elliptic flow appears to change so little over such a large 
range of beam energies, we investigate in more detail the contributions from the different 
phases of the heavy ion collision event on this observable. Figure~\ref{Figure_v2_phases} 
demonstrates the magnitude of $v_2$ before the hydrodynamical evolution, right after 
particlization and finally after the hadronic rescatterings performed in the UrQMD (the end 
result). In the most central collisions, where the overall elliptic flow is small compared 
to mid-central collisions, the effect of the hadronic rescatterings is negligible. In 
the impact parameter range $b = 8.2-9.4$ fm the contribution from the hadronic 
rescatterings is about 10\%. 

In both centralities, the hydrodynamic phase contributes very little to the elliptic flow
at $\sqrt{s_{NN}}=5 - 7.7$ GeV; $v_2$ at 5 GeV is in practice completely 
produced by the transport dynamics. However, already at $\sqrt{s_{NN}}=11.5$ GeV the 
contribution from the hydrodynamics is significant in non-central collisions. 
It thus seems that the hydrodynamically produced $v_2$ {\em does} vanish at low collision 
energies, as was the naive expectation. The measured $v_2$ still remains 
nonzero, however, as the transport dynamics become more important at lower energies and 
are able to compensate for the reduction of hydrodynamically produced flow. Indeed, a 
recent study by Denicol {\em et al.} suggests that the hadron resonance gas with a large 
baryon number density can have more ideal fluid-like behavior compared to the same gas at zero 
baryon number density \cite{Denicol:2013nua}.

\begin{figure}
\centering
\includegraphics[width=8cm]{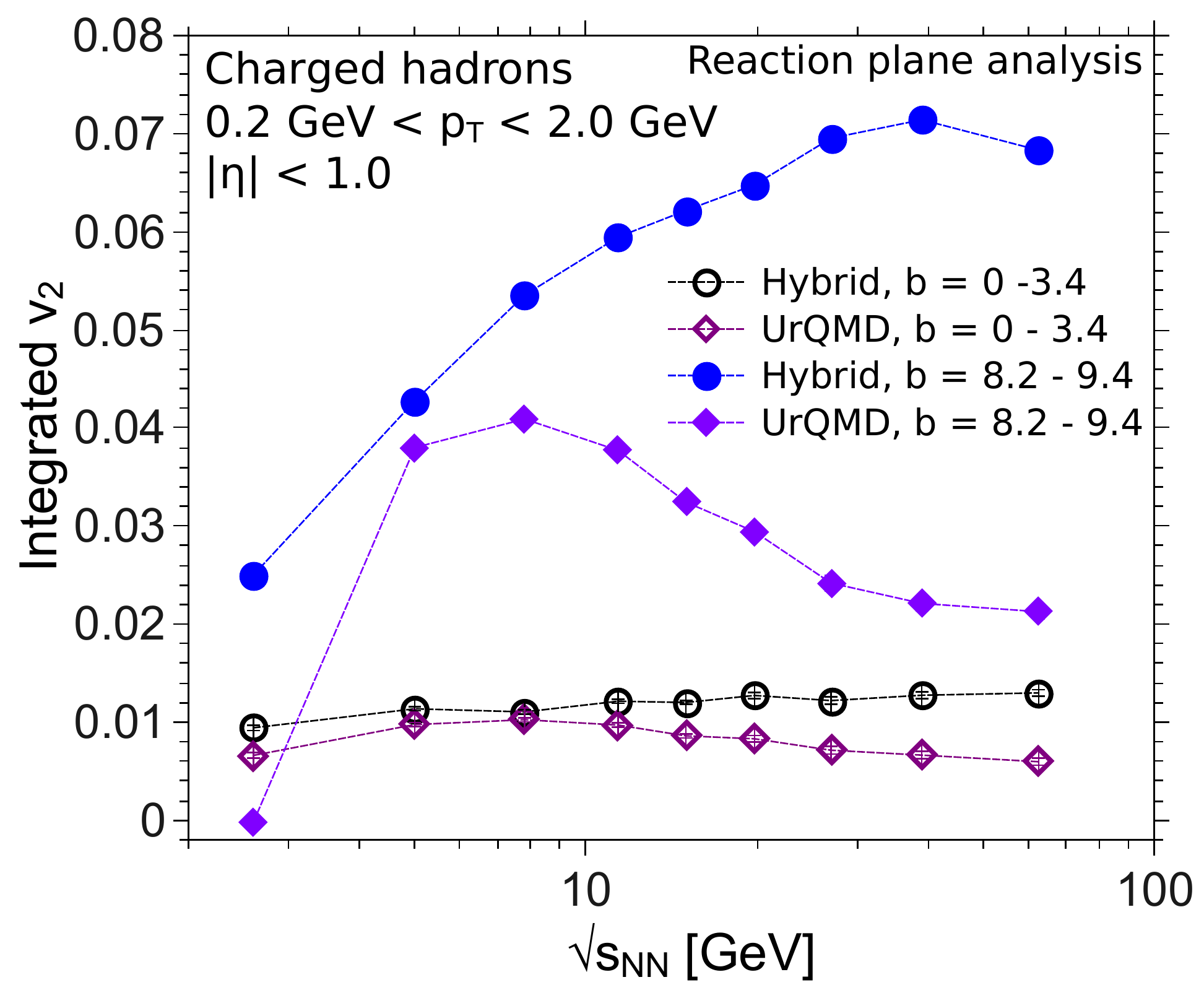}
\caption{(Color online) Comparison of $v_2\{\textrm{RP}\}$ produced in hybrid (circles) and 
in UrQMD without hydro (diamonds).}
\label{Figure_v2_rp}
\end{figure}

To make a connection with the earlier reaction plane analyses 
\cite{Petersen:2009vx,Petersen:2010md}, Figure~\ref{Figure_v2_rp} shows the integrated 
reaction plane $v_2$ for the present version of the hybrid, compared with the UrQMD result 
without hydrodynamics. This comparison also demonstrates that for $\sqrt{s_{NN}}=5$ GeV, 
the contribution from hydrodynamics is nearly negligible. At the even lower energy 
$E_{\textrm{lab}}=2$ GeV the hybrid again produces more flow; however, the applicability 
of ideal hydrodynamics for the whole system at such a low energy begins to be questionable. 
At lower energies a dynamical switching between non-equilibrium transport and fluid dynamics 
is necessary which is beyond the current capabilities of the hybrid model. 

\subsection{Triangular flow}
\label{sec_res_v3}

The triangular flow $v_3$ is a good observable for measuring the system sensitivity to 
the initial state fluctuations, as it is generated solely by event-by-event variations 
in the initial configuration of the colliding nucleons. Compared to the elliptic flow, 
triangular flow is considerably less sensitive to the overall collision geometry; it is, however, 
also harder to measure reliably due to having a smaller magnitude.

\begin{figure}
\centering
\includegraphics[width=7cm]{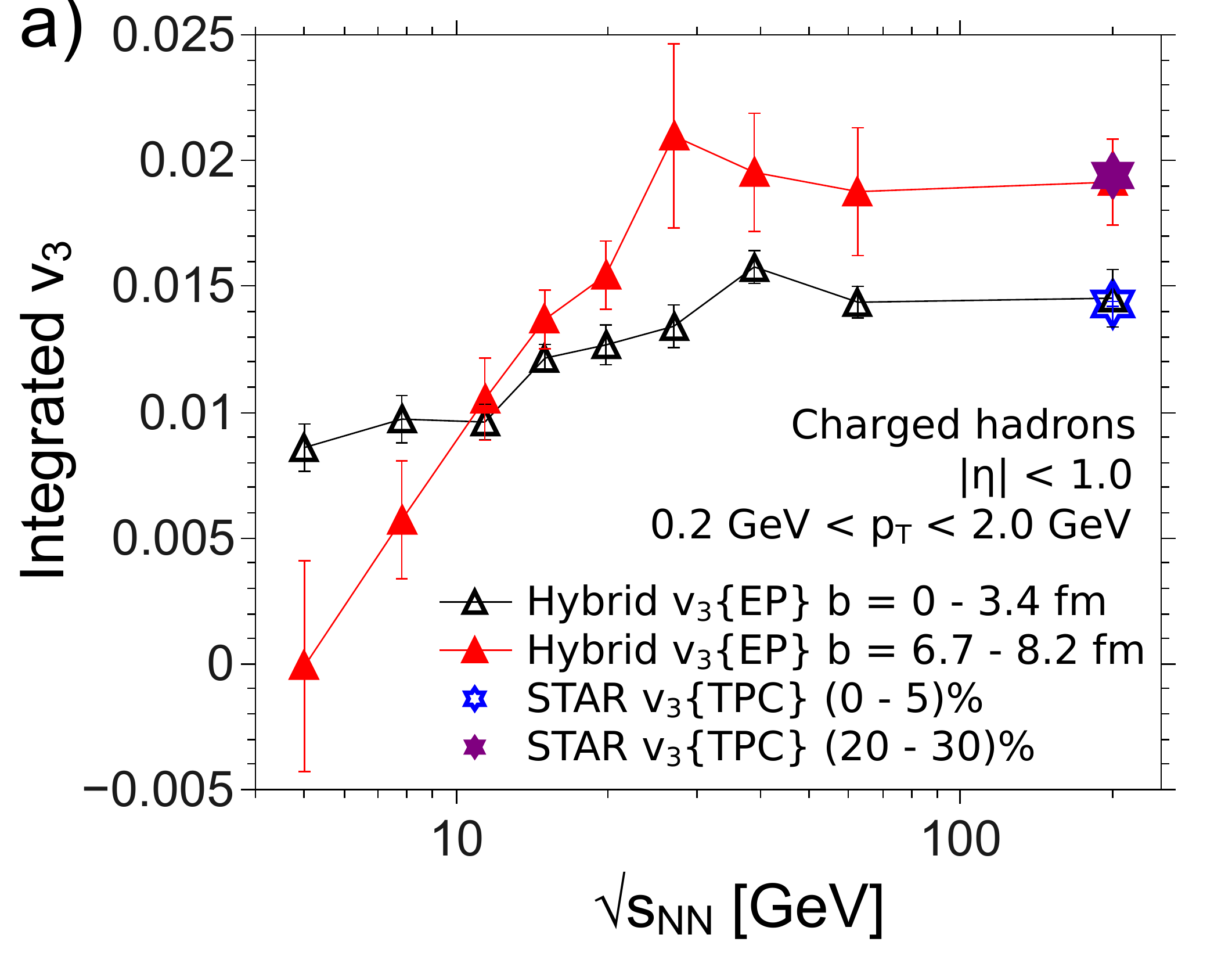}
\includegraphics[width=7cm]{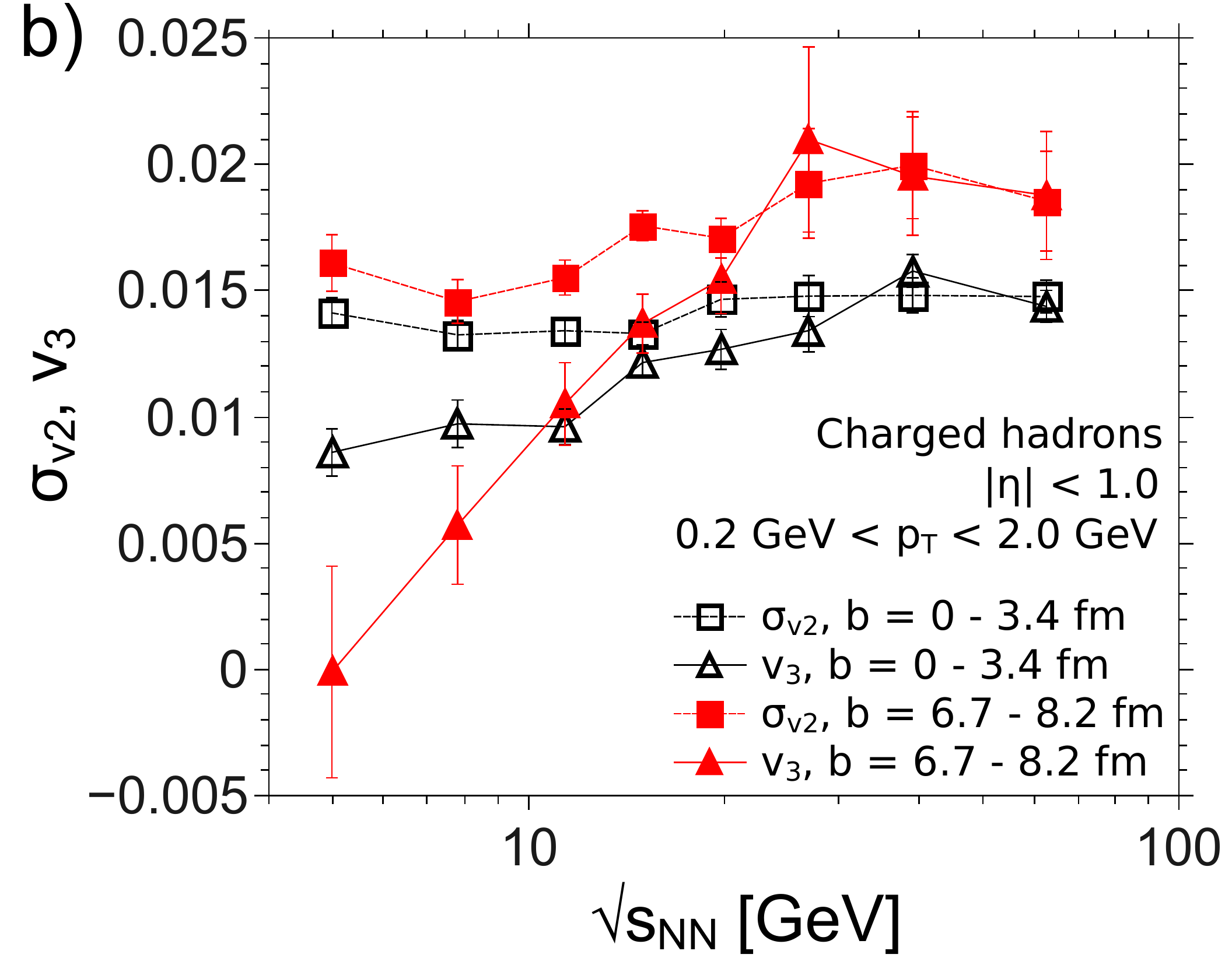}
\caption{(Color online) a) Integrated $v_3\{\textrm{EP}\}$ at midrapidity $|\eta|<1.0$ 
in central collisions ($b = 0-3.4$ fm, open triangles) and midcentral collisions 
($b = 6.7-8.2$ fm, solid triangles), compared with the STAR data \cite{Adamczyk:2013waa} (stars). 
b) $v_3\{\textrm{EP}\}$ compared with 
initial state fluctuations' contribution to $v_2$,
$\sigma_{v2}=\sqrt{\frac{1}{2}(v_2\{\textrm{EP}\}^2-v_2\{\textrm{RP}\}^2)}$ 
(squares).}
\label{Figure_v3}
\end{figure}

The $\sqrt{s_{NN}}$-dependence of the integrated $v_3\{\textrm{EP}\}$ produced by the 
hybrid model for impact parameter ranges $b = 0-3.4$ fm and $b = 6.7-8.2$ fm is presented 
in Figure~\ref{Figure_v3}a. In the most central collisions, $v_3$ increases only slightly 
from below 0.01 to 0.015 with increasing collision energy, whereas in midcentrality there 
is a rapid rise from $\approx 0$ at $\sqrt{s_{NN}} = 5$ GeV to the value of $\approx 0.02$ 
for $\sqrt{s_{NN}} \geq 27$ GeV. The values at $\sqrt{s_{NN}} = 200$ GeV are in agreement
with the experimental data \cite{Adamczyk:2013waa}. The energy dependence of midecentral 
$v_3$ is very similar to what was seen for the hydrodynamically produced $v_2$ in 
Figure~\ref{Figure_v2_phases}b, suggesting that in this case the transport part of the 
model is unable to compensate for the diminished hydro phase.

Like $v_3$, the event plane elliptic flow $v_2\{\textrm{EP}\}$ is also
affected by the initial state fluctuations, as the event plane angle (and thus the tilt
of the $v_2$-generating ellipsoid) varies event-by-event. 
The reaction plane $v_2$, on the other hand, is insensitive to these variations. 
Based on \cite{Bhalerao:2006tp,Voloshin:2007pc,Ollitrault:2009ie}, 
we (ignoring nonflow effects) define the contribution of fluctuations to $v_2$ as
\begin{equation}
\label{eq_v2fluc}
\sigma_{v2}=\sqrt{\frac{1}{2}(v_2\{\mbox{EP}\}^2-v_2\{\mbox{RP}\}^2)},
\end{equation}
and compare the magnitudes of the two fluctuation-based observables in 
Figure~\ref{Figure_v3}b. We find $\sigma_{v2}$ remaining nearly constant with respect to the 
collision energy. Within the statistical uncertainties, 
$v_3\{\textrm{EP}\}=\sigma_{v2}$ at $\sqrt{s_{NN}} \geq 27$ GeV. Thus the system is able to
convert the initial state fluctuations to $v_2$ at all energies, but for $v_3$ the task 
becomes increasingly more difficult with lower $\sqrt{s_{NN}}$.

\subsection{Flow dependence on initial geometry}
\label{sec_res_geometry}

Based on the above comparison of fluctuation-generated $\sigma_{v2}$ and $v_3$, the 
relationship between the flow coefficients and the initial collision geometry warrants more 
investigation.

\begin{figure}
\centering
\includegraphics[width=7cm]{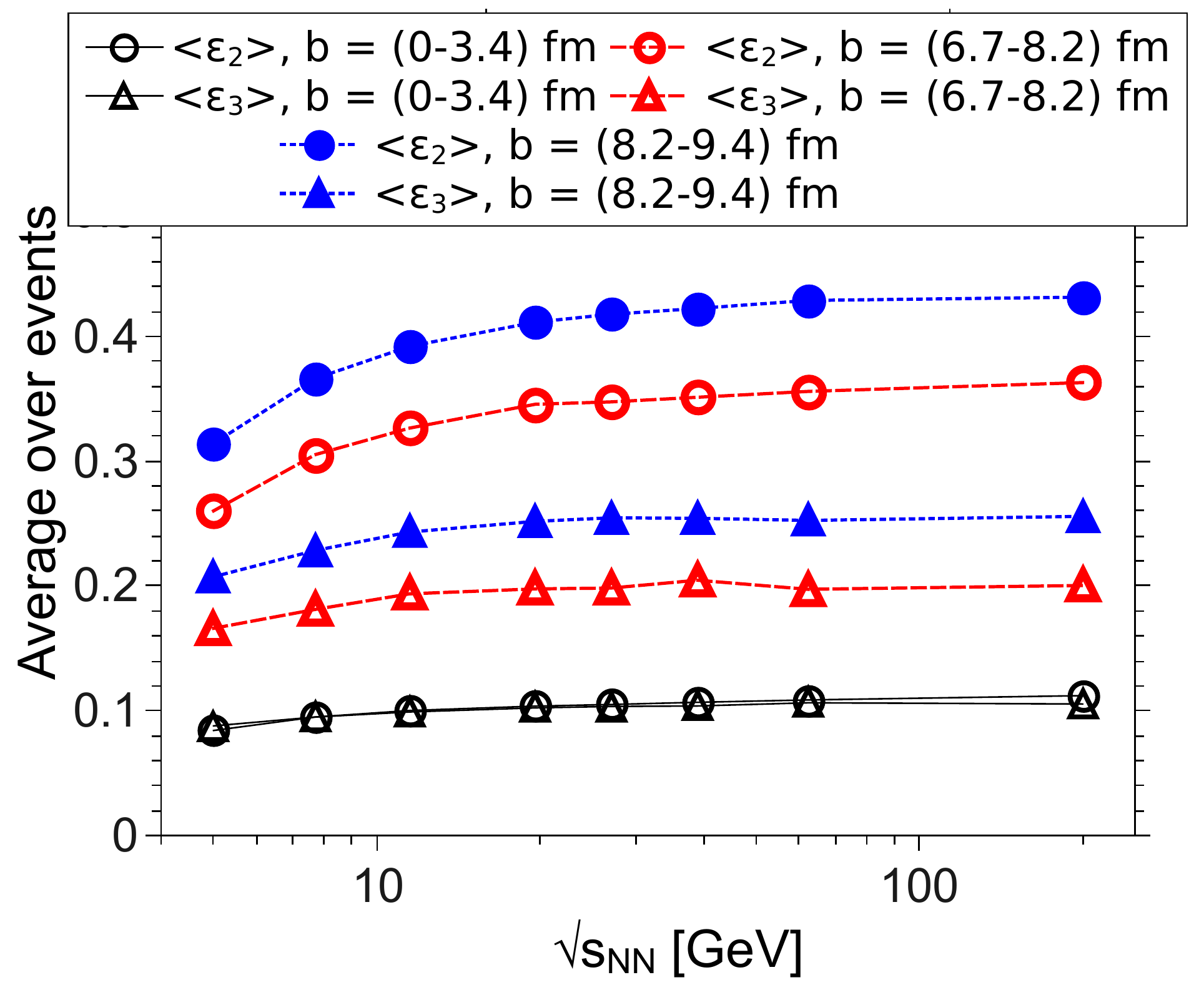}
\caption{(Color online) Average eccentricity $\langle \epsilon_2 \rangle$ (circles) and 
triangularity $\langle \epsilon_3 \rangle$ (triangles)
as a function of collision energy $\sqrt{s_{NN}}$, for impact parameter ranges
$b = 0-3.4$ fm (solid lines), $6.7-8.2$ fm (dashed lines) and $8.2-9.4$ fm (dotted lines).}
\label{Figure_eccentricity}
\end{figure}

Figure~\ref{Figure_eccentricity} illustrates the collision energy and centrality 
dependencies of the event-averaged initial state spatial eccentricity 
$\langle \epsilon_2 \rangle$ and triangularity $\langle \epsilon_3 \rangle$. 
The eccentricity and triangularity in an event, calculated at the beginning of 
the hydrodynamical evolution $t_{\textrm{start}}$, are defined by 
\cite{Petersen:2010cw}: 
\begin{equation}
\label{eq_eccentricity}
\epsilon_n=\frac{\sqrt{\langle r^n \cos(n\phi) \rangle^2+\langle r^n \sin(n\phi) \rangle^2}}{\langle r^n \rangle},
\end{equation}
where $(r,\phi)$ are the polar coordinates of the participant particles in the event 
and $\langle \dots \rangle$ denotes the average over the particles.

In the most central collisions, the collision area is nearly circular; both the average 
eccentricity and triangularity are created purely by the fluctuations in the spatial 
orientation of colliding nucleons and are similar in magnitude. At mid-central collisions, 
the overlap region of the colliding nuclei is almond-shaped, making 
$\langle \epsilon_2 \rangle$ clearly larger than $\langle \epsilon_3 \rangle$. As neither 
the typical spatial distribution of binary collisions, nor the inelastic nucleon-nucleon 
cross section $\sigma_{NN}$ change significantly within the examined energy range, one 
expects only a weak dependence on the collision energy. 

The hydro starting time $t_{\textrm{start}}$, however, is sensitive to 
the beam energy, dropping from 5.19 fm at $\sqrt{s_{NN}}=5$ GeV to 1.23 fm at 
$\sqrt{s_{NN}}=19.6$ GeV \cite{Petersen:2008dd}. Thus the main reason for the systematic 
decrease of  $\langle \epsilon_2 \rangle$ and $\langle \epsilon_3 \rangle$ seen
at low energies in Fig.~\ref{Figure_eccentricity} is the longer transport 
evolution before the start of the hydrodynamical phase. During this evolution finite $v_2$ 
and $v_3$ values are built up that quench the initial eccentricity and triangularity.  

\begin{figure}
\centering
\includegraphics[width=7cm]{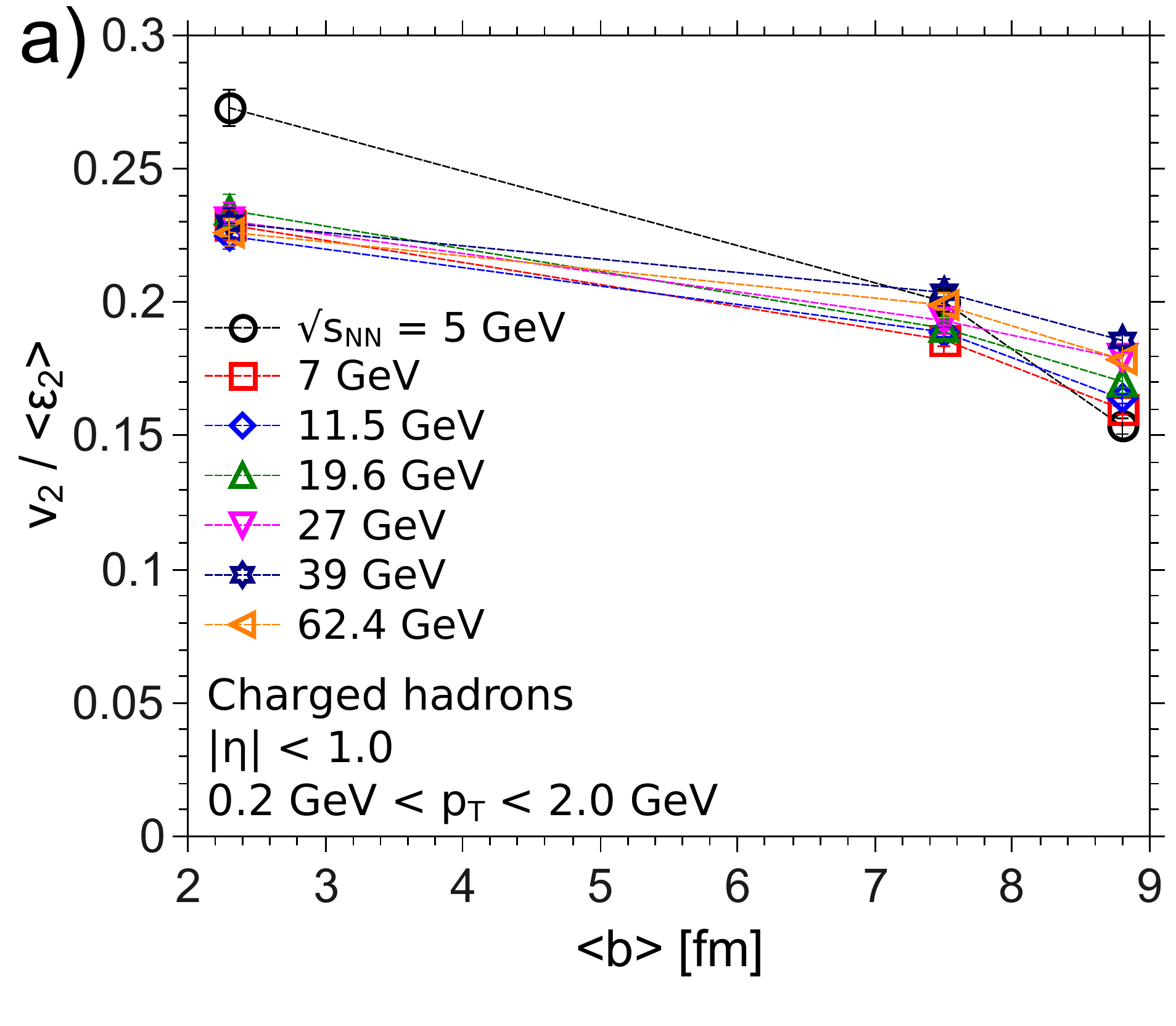}
\includegraphics[width=7.3cm]{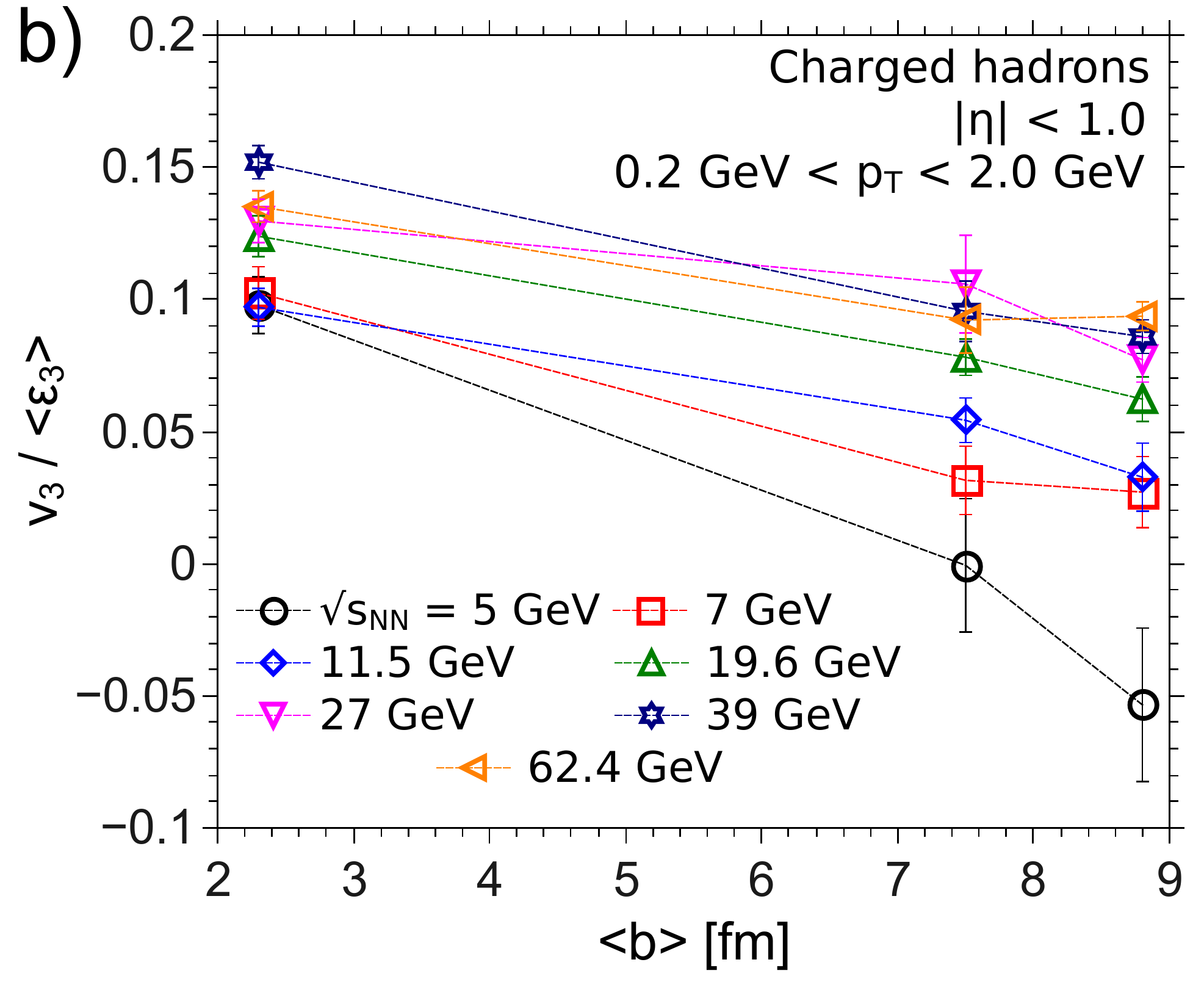}
\caption{(Color online) Energy evolution of a) $v_2\{\textrm{EP}\}$ scaled with average 
eccentricity $\langle \epsilon_2 \rangle$, and b) $v_3\{\textrm{EP}\}$ scaled with average 
triangularity $\langle \epsilon_3 \rangle$, as a function 
of average impact parameter $\langle b \rangle$.}
\label{Figure_vn_scaled_cent}
\end{figure}

In order to examine the system response to initial geometry, we scale $v_2$ and $v_3$ 
with $\langle \epsilon_2 \rangle$ and $\langle \epsilon_3 \rangle$, respectively. 
The result for the three centrality classes, represented by their average impact parameters 
$\langle b \rangle$, is shown in Figure~\ref{Figure_vn_scaled_cent}. 
Aside from the anomalous most central point at $\sqrt{s_{NN}}=5$ GeV, the relation of the 
elliptic flow to the initial eccentricity changes relatively little for the whole collision 
energy range, in comparison to the $v_3$ response to the triangularity of the initial state 
which shows a clear increase as one moves toward higher energies. This supports the idea 
that the hadron gas dynamics are sufficient for producing the $v_2$ response to 
the collision geometry at low collision energies, but a less viscous fluid would be needed 
for producing the comparatively weaker $v_3$ response to triangularity at the same 
$\sqrt{s_{NN}}$.

\begin{figure}
\centering
\includegraphics[width=6cm]{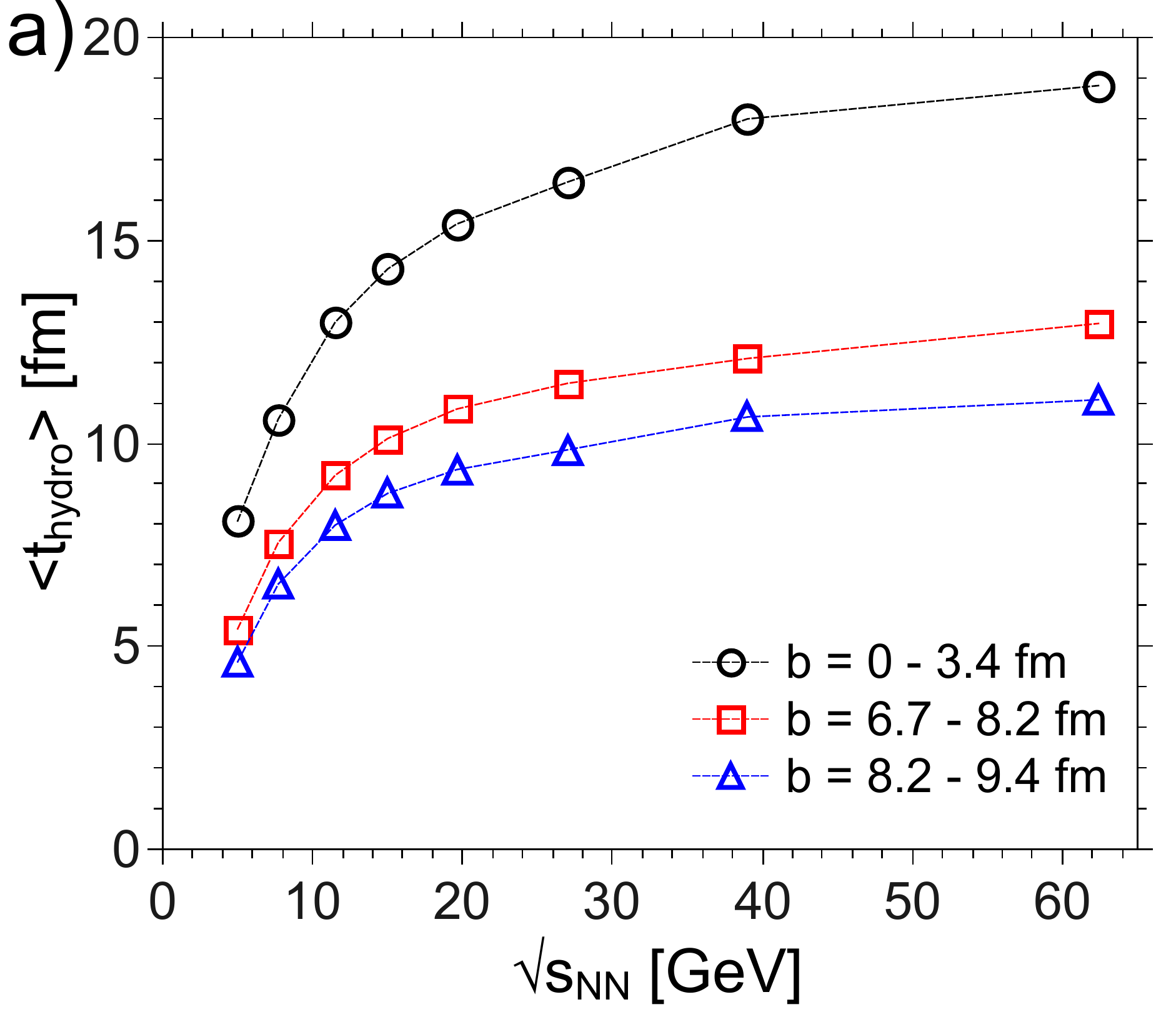}
\includegraphics[width=6.3cm]{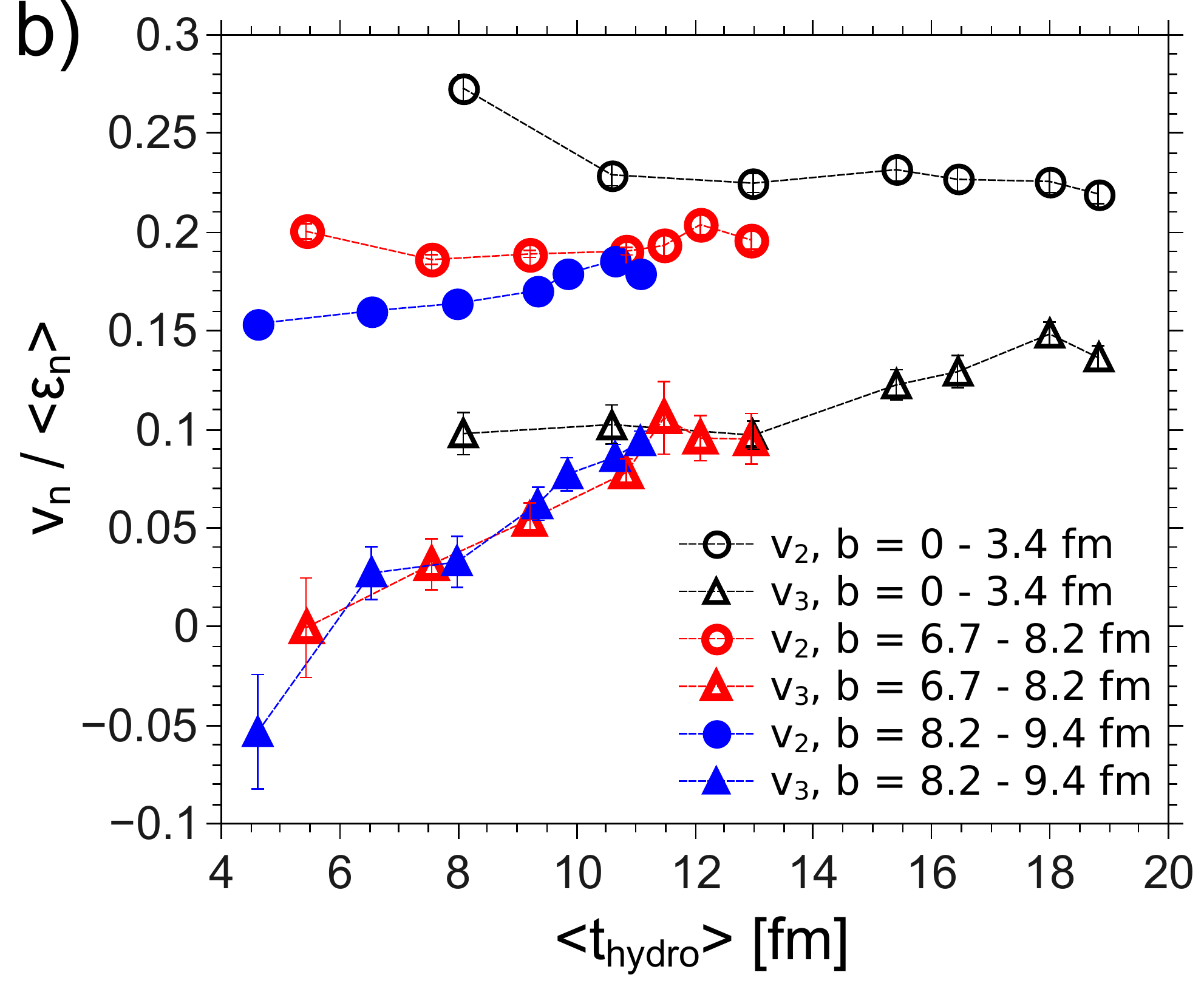}	
\caption{(Color online) a) Average total duration $\langle t_{\textrm{hydro}} \rangle$ 
of the hydrodynamical phase in the simulation as a function of collision energy 
$\sqrt{s_{NN}}$. b) Scaled flow coefficients $v_2/ \langle \epsilon_2 \rangle$ and 
$v_3/ \langle \epsilon_3 \rangle$ with respect to the 
average total hydro duration for impact parameter ranges $b = 0-3.4$ fm, $6.7-8.2$ fm 
and $8.2-9.4$ fm.}
\label{Figure_timedependence}
\end{figure}

To study the dependence of the flow coefficients on the existence of a hydrodynamic 
evolution in more detail, we plot, in Figure~\ref{Figure_timedependence}b, 
the scaled $v_2$ and $v_3$ with 
respect to the event-averaged total duration of hydrodynamical phase in the simulation 
$\langle t_{\textrm{hydro}} \rangle$, measured in the computational frame 
(Fig.~\ref{Figure_timedependence}a). It should be noted that this quantity represents 
the absolute upper limit of hydrodynamical phase in the simulation; as the particlization 
surface is not isochronous but iso-energy density, most of the system has been decoupled 
from the hydro long before $t_{\textrm{hydro}}$. For an example of the actual spacetime 
dependence of the particlization in the hybrid model, see Ref.~\cite{Huovinen:2012is}.

It is seen in Fig.~\ref{Figure_timedependence}b that, with the exception of the most central 
collisions at $\sqrt{s_{NN}}=5$ GeV, the scaled $v_3$ points form an uniform, monotonically 
increasing function of $\langle t_{\textrm{hydro}} \rangle$. For $v_2$, the different 
centralities do not have such uniform behavior because of the additional elliptic flow 
produced by the transport part at low energies. In other words, the final triangular flow 
is purely a product of hydrodynamics at all beam energies, while the final elliptic flow at 
low $\sqrt{s_{NN}}$ is not.

\section{Summary}
\label{sec_summary}

In this article, we have investigated the collision energy dependence of the flow 
coefficients $v_2$ and $v_3$ in a hybrid transport + hydrodynamics approach. 
In such a framework, it is seen that the hadron / string pre-equilibrium dynamics can 
compensate for the  diminished hydrodynamical evolution for $v_2$ production at lower 
collision energies. Because of this, $v_2$ changes relatively little as a function of 
beam energy. This remains true for $v_2$ scaled with the average eccentricity 
$\langle \epsilon_2 \rangle$, as the initial eccentricity also changes very modestly 
for the most of the examined collision energy range, decreasing more steeply only below 
$\sqrt{s_{NN}} \lesssim 10$ GeV, where the pre-equilibrium phase lasts for several fm.

For the triangular flow $v_3$, generated purely by the spatial configuration fluctuations of 
the colliding nucleons in the initial state, it is found that the system response to initial 
triangularity begins decreasing below $\sqrt{s_{NN}}=27$ GeV, reaching $\approx$ 0 for 
midcentral collisions at $\sqrt{s_{NN}}=5$ GeV. In addition, the scaled $v_3$ points over 
several collision energies and centrality classes form an uniform function of the hydro 
duration $\langle t_{\textrm{hydro}} \rangle$, whereas for the elliptic flow the relation is 
distorted by the transport dynamics. Thus, compared to $v_2$, the triangular flow provides 
a clearer signal for the formation of (near-)ideal fluid in heavy ion collisions.

For the future studies, the issues with kaon production and $v_2(p_T)$ overestimating 
the data at higher $p_T$ will necessitate a slight re-tuning of the model parameters and 
possibly the addition of viscous corrections to the hydrodynamical phase for the optimal 
agreement with the experimental data. Also, while the values for the triangular flow $v_3$ 
at high collision energies (and also at the lower limit $\sqrt{s_{NN}}=7.7$ GeV) 
quantitatively agree with the experimental results, there is a 
qualitative disagreement with the preliminary STAR data, where no beam energy dependence 
is seen within 0-5\%, 5-10\% or 10-20\% centrality at $\sqrt{s_{NN}}=7.7-27$ GeV 
\cite{Pandit:QM2012}. 

As the hadron resonance gas has proven to be too viscous for producing $v_3$ from initial 
state fluctuations in this investigation, the current discrepancy between the simulation 
results and the preliminary experimental data implies that larger quantities of low-viscous 
state of matter is manifested at lower collision energies than expected in this study. 
On the other hand, the inconsistent behavior of flow observables at $\sqrt{s_{NN}}=5$ GeV 
compared to the higher energy points suggest that the lower energy limit of 
applicability may have been reached for the ideal hydrodynamics approach. More detailed 
studies, both theoretical and experimental, are thus needed for $v_3$ at
$\sqrt{s_{NN}}\leq 10$ GeV energies.

\section{Acknowledgements}

We thank H.~Holopainen, Iu.~A.~Karpenko and P.~Huovinen for useful discussions. 
The authors acknowledge funding of the Helmholtz Young Investigator Group VH-NG-822. 
This work was supported by the Helmholtz International Center for the Facility for 
Antiproton and Ion Research (HIC for FAIR) within the framework of the 
Landes-Offensive zur Entwicklung Wissenschaftlich-\"okonomischer Exzellenz
(LOEWE) program launched by the State of Hesse. Computational resources have been provided 
by the Center for Scientific Computing (CSC) at the Goethe-University of Frankfurt.

\end{document}